\documentclass[runningheads]{llncs}
\usepackage{graphicx}
\usepackage{algorithmic}
\usepackage[ruled,vlined]{algorithm2e}
\usepackage{url}
\usepackage{color}

\usepackage{times}
\usepackage{textcomp}

\usepackage{amsmath}
\usepackage{amssymb}

\usepackage{fancyhdr}
\usepackage{color}
\usepackage{subfigure}
\usepackage{listings}

\usepackage{multicol}
\usepackage{balance}

\usepackage{booktabs}
\usepackage{pbox}
\usepackage[colorlinks]{hyperref}

\hypersetup{citecolor=blue, linkcolor=blue, urlcolor=blue}

\newcommand{\ida}{{iDataAgent}\xspace}

\newcommand{\sysname}{{PrivacyGuard}\xspace}

\newcommand{\algfootnote}{\footnotesize}

\SetKwInput{KwCEInit}{Initialization}
\SetKwInput{KwCECompute}{Compute}
\SetKwInput{KwCEUseRecord}{Usage Recording}
\SetKwProg{Fn}{Function}{}{}

\newcommand{\mdb}{{DB}\xspace}
\newcommand{\mdo}{{DO}\xspace}
\newcommand{\mdos}{{DOs}\xspace}
\newcommand{\mdc}{{DC}\xspace}

\newcommand{\mida}{{iDA}\xspace}
\newcommand{\tcee}{CEE\xspace}
\newcommand{\ctdo}{$\mathbf{C_{DO}}$\xspace}
\newcommand{\ctdb}{$\mathbf{C_{DB}}$\xspace}

\begin{document}
%
\title{PrivacyGuard: Enforcing Private Data Usage Control with Blockchain and Attested Off-chain Contract Execution}
\titlerunning{PrivacyGuard: Enforcing Private Data Usage Control}
%
\author{Yang Xiao\inst{1}\thanks{To present in the 25th European Symposium on Research in Computer Security (ESORICS 2020), September 14 - 18, 2020.} \and
Ning Zhang\inst{2} \and
Jin Li\inst{3} \and
Wenjing Lou\inst{1} \and
Y. Thomas Hou\inst{1} 
}
\authorrunning{Y. Xiao et al.}
%
\institute{Virginia Polytechnic Institute and State University, VA, USA \and
Washington University in St. Louis, MO, USA \and
Guangzhou University, Guangzhou, China
}

\maketitle              

\begin{abstract}
The abundance and rich varieties of data are enabling many transformative applications of big data analytics that have profound societal impacts. However, there are also increasing concerns regarding the improper use of individual data owner's private data. In this paper, we propose PrivacyGuard, a system that leverages blockchain smart contract and trusted execution environment (TEE) to enable individual's control over the access and usage of their private data. Smart contracts are used to specify data usage policy, i.e., who can use what data under which conditions and what analytics to perform, while the distributed blockchain ledger is used to keep an irreversible and non-repudiable data usage record. To address the efficiency problem of on-chain contract execution and to prevent exposing private data on the publicly viewable blockchain, PrivacyGuard incorporates a novel TEE-based off-chain contract execution engine
along with a protocol to securely commit the execution result onto blockchain. 
We have built and deployed a prototype of \sysname with Ethereum and Intel SGX. Our experiment result demonstrates that \sysname fulfills the promised privacy goal and supports analytics on data from a considerable number of data owners. 

\end{abstract}
%
%
%

\section{Introduction}

The recent emergence of big data analytics and artificial intelligence has made life-impacting changes in many sectors of society. 
One of the fundamental enabling components for the recent advancements in artificial intelligence is the abundance of data. However, as more information on individuals is collected, shared, and analyzed, there is an increasing concern on the privacy implication. In the 2018 Facebook-Cambridge Analytica data scandal, an API, originally designed to allow a third party app to access the personality profile of limited participating users, was misused by Cambridge Analytica to collect information on 87 million of Facebook profiles without the consent of the users. These illicitly harvested private data were later used to create personalized psychology profiles for political purposes~\cite{facebookScandalWeb}. With increasing exposure to the privacy risks of big data, many now consider the involuntary collection of personal information a step backward in the fundamental civil right of privacy \cite{eu2016gdpr}, or even in humanity~\cite{tedTalkCayla,timcookTalk}. Yet, driven by economic incentives, the collection and analysis of the personal data continue to grow at an amazing pace. 

Individuals share personal information with people or organizations within a particular community for specific purposes; this is often referred to as the context of privacy~\cite{nissenbaum2004privacy}. For example, individuals may share their medical status with healthcare professionals, product preferences with retailers, and real-time whereabouts with their loved ones. When information shared within one context is exposed in another unintended one, people may feel a sense of privacy violation~\cite{NPRS2016}. The purposes and values of those contexts are also undermined. The contextual nature of privacy implies that privacy protection techniques need to address at least two aspects: 1) what kind of information can be exposed to whom, under what conditions; and 2) what is the ``intended purpose" or ``expected use" of this information. 

Much research has been done to address the first privacy aspect, focusing on data access control \cite{goyal2006attribute,bethencourt2007ciphertext,wang2010hierarchical,yu2010achieving} and  
data anonymization \cite{dwork2008differential,machanavajjhala2007diversity,sweeney2002k,li2007t}. Only recently, there have been a few works that attempted to address the second aspect of privacy from the architecture perspective~\cite{Zyskind2015SPW,zyskind2015enigma,sen2014bootstrapping,elnikety2016thoth,datta2017use,intelPDO}. In fact, many believe that the prevention of this kind of ``second-hand" data (mis)use can only be enforced by legal methods~\cite{datareuseTaxonomyYaleLaw}. Under the current practice, once an authorized user gains access to the data, there is little control over how this user would use the data. Whether he would use the data for purposes not consented by the original data owner, or pass the data to another party (i.e., data monetization) is entirely up to this new ``data owner'', and is no longer enforceable by the original data owner. 

\textbf{Our Contribution~}
Building upon our previous work \cite{zhang2018privacyguard}, we present the design, implementation and evaluation of \emph{\sysname} in this paper. \sysname empowers individuals with full control over the access and usage of their private data in a data market. The data owner is not only able to control who can have access to their private data, but also ensured that the data are used only for the intended purpose. To realize this envisioned functionalities of \sysname, three key requirements need to be met. First, users should be able to define their own data access and usage policy in terms of to whom they will share the data, at what price, and for what purpose. Second, data usage should be recorded in a platform that offers non-repudiation. Third, the actual usage of data should have a verifiable proof to show its compliance to the policy. 

Blockchain, the technology behind Bitcoin \cite{nakamoto2008bitcoin} and Ethereum \cite{EthereumWeb}, has exhibited great potential in providing security and privacy services.
Smart contract is a program that realizes a global state machine atop the blockchain and has its correct execution enforced by the blockchain's consensus protocol. \sysname enables individual users to control the access and usage of their data via smart contract and leverages the blockchain ledger for transparent and tamper-proof recording of data usage.

While smart contract and blockchain appear to be the perfect solution, there are fundamental limitations if applied directly. First, data used by smart contracts are uploaded in the form of blockchain transaction payload, which is not designed to hold arbitrarily large amount of data due to communication burden and scalability concerns \cite{nakamoto2008bitcoin,croman2016scaling}.
Second, smart contracts are small programs that have to be executed by all participants in the network, which raises serious computational efficiency concerns. For the same reason, existing platforms such as Ethereum are not purposed to handle complex contract programs \cite{wust2019ace}.
Last but not least, data used by smart contracts are available to every participant on blockchain by design, which conflicts with the confidentiality requirement of user data. Existing secure computation techniques for preserving confidentiality and utility of data, such as functional encryption~\cite{boneh2011functional}, can nonetheless be prohibitively expensive for the network.

To tackle data and computation scalability problems, 
\sysname splits the private data usage enforcement problem into two domains: the control plane and the data plane. In the control plane, individual users publish the availability as well as the usage policy of their private data as smart contracts on blockchain. 
Data consumers interact with the smart contract to obtain authorization to use the data. Crucially, the actual data of the users are never exposed on the blockchain. Instead, they are stored in the cloud in encrypted forms. Computation on those private user data as well as the provision of secret keys are accomplished off-chain in the data plane with a trusted execution environment (TEE)~\cite{sgx,trustzone} on the cloud. 
%

%
When a data contract's execution is split into control and computation, where the computation actually takes place off-chain, several challenges occur. First, the correctness of the contract execution can no longer be guaranteed by the blockchain consensus. To this end, we propose ``local consensus'' for the contracting parties to establish trust on the off-chain computation via remote attestations. Second, the execution of contract is no longer atomic when the computation part is executed off-chain. We design a multi-step commitment protocol to ensure that result release and data transaction remain an atomic operation, where if the computation results were tampered with, the data transaction would abort gracefully. Lastly, private data are protected inside the TEE enclave and secrets are only provisioned when approved according to the contract binding.



We implemented a prototype of \sysname using Intel SGX as the TEE technology and Ethereum as the smart contract platform. 
We chose these two technologies for implementation due to their wide adoption.
Our design generally applies to other types of trusted execution environments and blockchain smart contract platforms. 
The platform fulfills the goal of user-define data usage control at reasonable cost and we show that it is feasible to perform complex data operations with the security and privacy protection as specified by the data contract.

To summarize, we make the following contributions in this paper:
\begin{itemize}
\item We propose \sysname, a platform that combines blockchain smart contract and trusted execution environment to address one of the most pressing problems in big data analytics---trustworthy private data computation and usage control.
\sysname essentially allows data owners to contribute their data into the data market and specify the context under which their data can be used.

\item We propose a novel construction of off-chain contract execution environment to support the vision of \sysname, which is the key to improving the execution efficiency of smart contract technology and enabling trustworthy execution of complex contract program without solely relying on costly network consensus.

\item We implemented a prototype of \sysname using Intel SGX and Ethereum smart contract and and deployed it in a simulated data market. Our evaluation shows that \sysname is capable of processing considerable volumes of data transactions on existing public blockchain infrastructure with reasonable cost.

\end{itemize}

\section{Background}

\textbf{Blockchain and Smart Contract~}
Blockchain is a recently emerged technology used in popular cryptocurrencies such as Bitcoin~\cite{nakamoto2008bitcoin} and Ethereum~\cite{EthereumWeb}. It enables a wide range of distributed applications as a powerful primitive. With a blockchain in place, applications that could previously run only through a trusted intermediary can now operate in a fully decentralized fashion and achieve the same functionality with comparable reliability. When the majority of the network's voting power (hashing power, stake value, etc.) are controlled by honest participants, the shared blockchain becomes a safe and timestamped record of the network's activities.
The conceptual idea of programmable electronic ``smart contracts'' dates back nearly twenty years~\cite{szabo1997formalizing}. When implemented in the blockchain platform (eg. Ethereum), smart contracts are account-like entities that can receive transfers, make decisions, store data or even interact with other contracts. 
The blockchain and the smart contract platform however have several drawbacks in transaction capacity~\cite{croman2016scaling}, computation cost~\cite{ekiden,kalodner2018arbitrum}, as well as privacy of user and data~\cite{hawk,kalodner2018arbitrum}. 

\noindent
\textbf{Trusted Execution Environment~}
Creating vulnerability-free software has long been considered a very challenging problem~\cite{song2019sok}.
Researchers in the architecture community in both academia~\cite{costan2016sanctum} and industry~\cite{trustzone,sgx} have embraced a new paradigm of limiting the trusted computing base (TCB) to only the hardware, realizing a trusted execution environment (TEE). The well-known Intel SGX \cite{sgx} is an instruction set extension to provide TEE functionalities. Applications are executed in secure containers called \textit{enclaves}. The hardware guarantees the integrity and confidentiality of the protected application, even if the platform software is compromised. 
TEE has recently been adapted as a powerful tool to support blockchain-based applications~\cite{fisch2017iron,zhang2016town,ryoan,ekiden}.



\section{\sysname Overview}


\subsection{System Goal and Architecture}
The vision of \sysname is to not only protect data owner privacy but also promote a vibrant data sharing economy, in which data owners can confidently sell the right to use their data to data consumers for profits without worrying about data misuse. To realize this, there are three specific goals.
    First, data encryption/decryption are fully controlled by data owners. Untrusted parties (eg. cloud storage and data consumers) can not obtain or possess data owners’ plaintext data.
    %
    Second, Data owners are able to control who can access which data items under what conditions for what usage. The data usage records should be non-repudiable and auditable by data owners.
    %
    Third, the security mechanism of our system should be able to capture user-defined policies and enforce the compliance of the policies during the execution of data access.

\begin{figure}
    \vspace{-0.1in}
    \centering
    \includegraphics[width=4.5in]{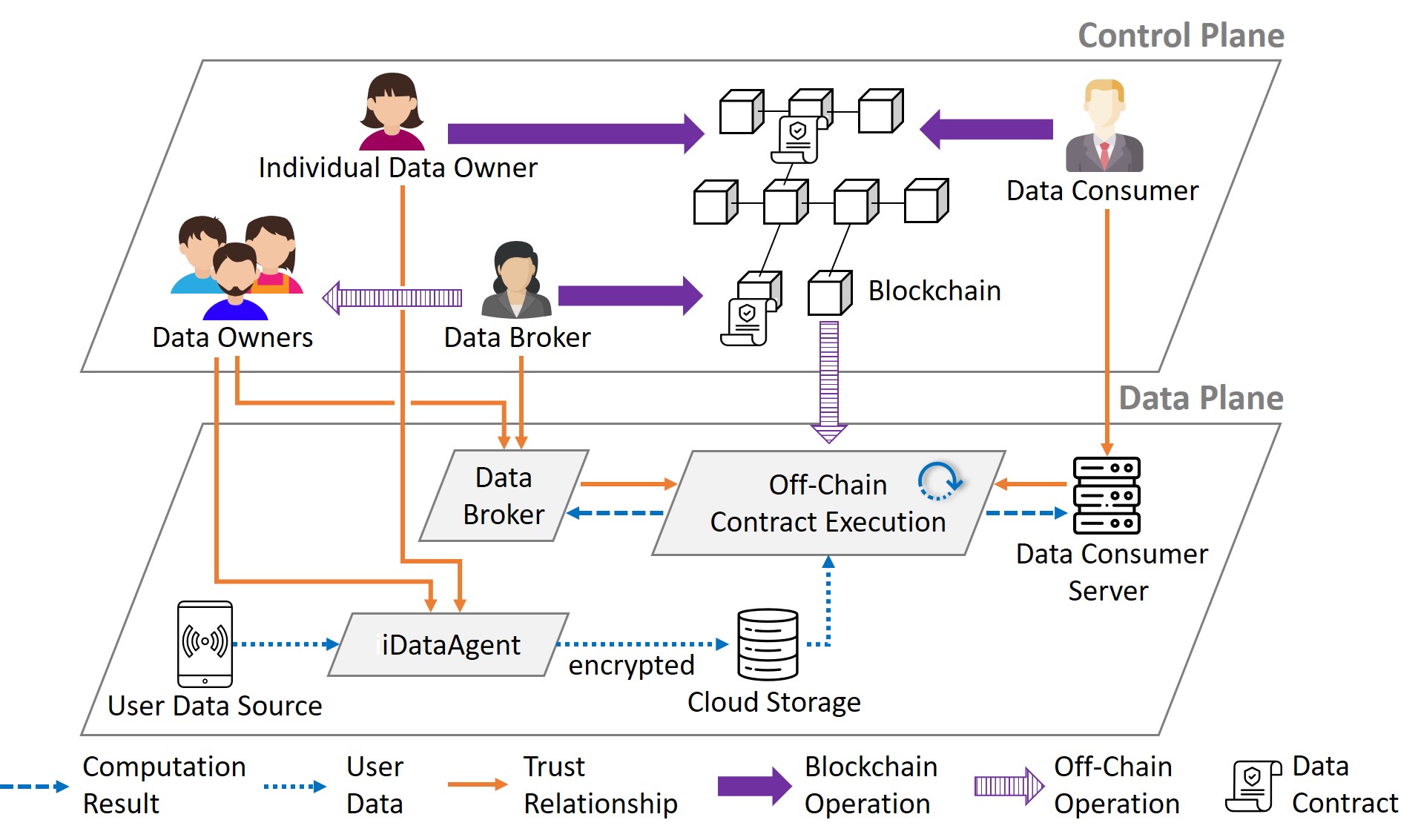}
    \caption{System architecture for \sysname framework.}
    \label{fig.Arch}
    \vspace{-0.1in}
\end{figure}
 


Fig.~\ref{fig.Arch} shows the system architecture of \sysname. Although we have been using the term \emph{users} to refer to both individuals and organizations, we differentiate two roles that an user in the data market can take. 
We refer to the individual or organization that owns the data as \emph{data owner (\mdo)} and the entity that needs to access the data as \emph{data consumer (\mdc)}.
Classified by the assigned responsibility, there are three main functional components in the PrivacyGuard framework: 

\textbf{Data Market~} Data market is an essential \sysname subsystem that supports the supply, demand and exchange of data on top of blockchain. For data access and usage control, \mdo can encode the terms and conditions pertaining to her personal data in a data contract, and publish it on a blockchain platform such as Ethereum. Data usage by \mdc is recorded via transactions that interact with the data contract. 
%


\textbf{\ida (\mida) and Data Broker (\mdb)~} \mida is a trusted entity representing an individual \mdo and responsible for key management for the \mdo. It also participates in contract execution by only provisioning the data key material to attested remote entities. 
Since it is often not realistic to expect individual \mdo to be connected all the time, \mida can also be instantiated as a trusted program in a TEE-enabled cloud server. To address the inherent transaction bandwidth limit of the blockchain network, \mdb is introduced to collectively represent a group of users.

\textbf{Off-chain Contract Execution Environment (CEE)~} This off-chain component executes data operations contracted between DO(s) and \mdc in a TEE enclave. The trusted execution guarantees correctness as if it was executed on-chain. The computation result is securely committed to \mdc while enforcing the contract obligation.

\subsection{\sysname High-Level Workflow} 
The workflow of \sysname proceeds in three stages which can function concurrently. Stage 1 and 2 involve the supply side (\mdo, \mida, \mdb) that prepares the data items and usage contracts while stage 3 characterizes the regular operation.

\textbf{Stage 1: Data Generation, Encryption, and Key Management~}
In this stage, a \mdo's data are generated by its data sources and collected by its \mida, who passes the encrypted data to the cloud storage. Keys for data en/decryption are generated by the \mdo via interface to \mida and managed by \mida. For a group of users with common data types, they can delegate their trust to a \mdb by remote-attesting the \mdb's enclave and provision data keys to the enclave.

\textbf{Stage 2: Policy Generation with Smart Contract~}
In \sysname, individual \mdos can define their own usage policies for their private data in \mdo contract (\ctdo). The policies encoded usually includes the essential components for privacy context, such as \emph{data type, data range, operation, cost, consumer, expiration}, etc. The operation, which specifies intended usage of the targeted data, can be an arbitrary attestable computer program. 
This paradigm grants \mdos fine-grained control on the data usage policy and the opportunity to participate in the data market independently. However, it requires ample transaction processing capacity from the blockchain network that scales in the number of \mdos.
Alternatively, the \mdb-based paradigm uses \mdb as a trusted delegate for a large number of \mdos. \mdb represents the \mdos in the blockchain by curating a \mdb contract (\ctdb) that accepts data registries from \mdos and advertising their data in bundles. The encoding of \ctdo and \ctdb will be elaborated in Section \ref{sec:controlPlane}.

\textbf{Stage 3: Data Utilization and Contract Execution~}
%
%
%
%
\mdc invokes a \ctdo (or \ctdb) for permission to use certain private data of the targeted DO(s) for a specific operation, and deposits payment onto the contract. If permission is granted on the blockchain, \mdc instructs \tcee to load the enclave program for the contracted operation whose checksum is specified in the contract. Then both the \mdc and \mida (or \mdb) proceed to remote-attest the \tcee enclave. This essentially allows the two parties to reach a ``local consensus'' on \tcee's trustworthiness that enables the \emph{off-chain execution} of the on-chain contract.
When the attestations succeed, \mida (or \mdb) provisions data decryption keys to the \tcee enclave to enable data operation within the enclave. When the operation finishes, the enclave releases the result in encrypted form and erases all the associated data and keying materials. To achieve a fair and atomic exchange that \mdc gets the decrypted result while DO(s) get the payment,
we propose a commitment protocol for the two sides which ensures the atomic exchange only when they agree upon each other. The detailed design of the commitment protocol will be explained in Section~\ref{sec:offchainexe}.

\subsection{Threat Model and Assumptions}
\label{subsec:threat-model}
We assume all entities act based on self-interest and may not follow the protocol. However, to maintain a reasonable scope for the paper, we assume \mdo will not provide meaningless or falsified data intentionally. It is possible to encode rules in smart contract to penalize \mdos for abusing the system with bad data. 
Furthermore, we assume the security systems, i.e. the blockchain and TEE, are trustworthy and are free of vulnerability. 
Specifically, in the control plane, we assume the blockchain infrastructure is secure that adversaries do not control enough resources to disrupt distributed consensus. We also assume smart contract implementations are free of software vulnerability.
%
%
In the data plane, we assume the TEE is up to date, and particularly, Intel SGX, is secure against malicious attack from the operating system. We recognize that TEE implementations are not always perfect, and previous work has demonstrated side channel information leakage on the SGX platform alone~\cite{van18foreshadow,wang2017leaky,zhang2016cachekit,zhang2018trusense,van2017telling,xu2015controlled,zhang2016case},
preventing such attacks is an important but orthogonal task. We also assume that all data operations requested by \mdc 
have been ratified by trusted sources and a cryptographic checksum of the program binary is sufficient for \sysname to check the data operation integrity.



\section{Data Market of User-Defined Usage with Blockchain}
\label{sec:controlPlane}



The intuition behind the data market is to enable fair and transparent data transactions between \mdo and \mdc. In \sysname, \mdos advertise private data items available for knowledge extraction on blockchain smart contracts. \mdc shops for a desirable data set and contract for his analytics. To start the data transaction, \mdc invokes the data contract and deposits a payment. The sales of knowledge extraction rights on private data are fulfilled that \mdo obtains the payment while the \mdc obtains the knowledge. The data transaction is then recorded in the blockchain with transparency.
To enable user-defined access and usage control, the data contract, needs to encode \mdo's data usage policy including how data can be used by which \mdc at what cost. Next we present the our data contract design in \sysname in a constructive manner.



\subsection{Encoding Data Usage Policy with Smart Contract}

%
%

%

\textbf{Basic Data Usage Contract~} In the conventional data sharing scenario, the data access policy often includes attributes such as type of the data, range or repository of the data, \mdo and \mdc credentials. For example, we assume patient $X$ with public key pair ($pk_{X}$,$sk_{X}$) has three types of medical data: radiology data, blood test data and mental record data. $X$ is only willing to share his radiology data (with descriptor $pData$) with urology specialist $S$ with public key $pk_{S}$. $X$ can treat $S$ as a \mdc and specify an access policy $P$ in a \emph{data access} contract: $\mathbf{C_{X(DA)}} = \{P=\{pData, pk_{S}\}, Sig_{sk_{X}}(P)\}$. This encoding, however, specifies only data access but no obligation of the \mdc once access is granted. The \mdc could share the data with other parties against the original intention of the \mdo. To enable fine-grained control on how data is used, obligations need to be attached to the policy. For instance, if $X$ only wants $S$ to run a certain operation $op$ on the data, then $X$ can encode a new \emph{data usage} contract in the following form: $\mathbf{C_{X(DU)}} = \{P=\{pData, op, pk_{S}\}, Sig_{sk_{X}}(P)\}$.

\textbf{Enabling Data Market Economy~}
A key feature of \sysname is to encourage \mdos to share private data for public welfare as well as financial rewards without concerning privacy leakage or data misuse. Building on top of the success of cryptocurrency, the blockchain smart contract platform allows \mdo and \mdc to transact on the usage of data with financial value attached.
%
\mdo can specify a price tag $\$pr$ (in cryptocurrency) in the policy. To further ensure a fair exchange that \mdc gets the knowledge and \mdo gets the payment, certain control logic should be instated in the form of smart contract functions. We call these functions and other contract metadata the contextual information, denoted $ctx$.
Back to the previous example, we now have $\mathbf{C_{X(DU)}}=\{P=\{pData, op, \$pr, pk_{S}\},ctx, Sig_{sk_{X}}(P||ctx)\} $. In blockchain domain, the signature is conveniently fulfilled by $X$'s signature in the contract creation transaction.

\textbf{Transparent Tracking of Data Utilization~}
For the system to provide transparent data utilization tracking and policy compliance auditing, each data transaction needs to be recorded in a tamper-resistant and non-repudiable manner. In \sysname, contract functions (part of $ctx$, invoked via blockchain transactions) are used to facilitate the recording of data utilization. Since the blockchain ledger is publicly managed via global consensus and unforgeable, contract function invocations in blockchain transactions can provide non-repudiable records on data utilization. 

\SetKwInput{KwInit}{Init}
\SetKwInput{KwRequestDataUse}{RequestDataUse}
\SetKwInput{KwRequest}{Request}
\SetKwInput{KwFreeze}{Freeze}
\SetKwInput{KwCompute}{Compute}
\SetKwInput{KwFinalize}{Finalize}
\SetKwInput{KwCancel}{Cancel}
\SetKwInput{KwRecord}{Record}
\SetKwInput{KwRevokeContract}{RevokeContract}
\SetKwInput{KwRevoke}{Revoke}
\SetKwInput{KwCompComplete}{ComputationComplete}
\SetKwInput{KwCompleteTrans}{CompleteTransaction}

\begin{algorithm}[H]
    \algfootnote
	\caption{Data Owner's Smart Contract \ctdo Pseudocode}
	\label{alg.dataContract}
    \scriptsize

\Fn{Constructor() ~~~~ \texttt{// Contract creation by \mdo with a policy}}{
	Parse $policy$ as ($dataset, price, operation, DCList, requestTimeout$) \;
	$pDS 				\leftarrow policy.dataset$ 			\;
	$pPrice 				    \leftarrow policy.price$ 				\;
	$pOP 	            \leftarrow policy.operation$ 			\;
	$pDCL              	\leftarrow policy.DCList$ 	        \;
	$pRTO             \leftarrow policy.requestTimeout$            \;
    $R \leftarrow [~]$  ~~~~ \texttt{// Usage records} \;
	$DO \leftarrow creator$ \;
}

\Fn{Request($op,data,\$f$) ~~~~ \texttt{// Callable by \mdc}}{
    \eIf{$op=pOP$ \textbf{and} $sender\in pDCL$ \textbf{and} $data\subset pDS$ \textbf{and} $f\geq pPrice $ }
	{  
	    Create a record entry $R[idx]$ with index $idx$ for this new data transaction \;
	    $R[idx].\{data,DC,reqTime\}\leftarrow \{data,sender,sys.time\}$ \;
	    $R[idx].status\leftarrow\textsc{wait\_computation}$ \;
	}
    {
        Return $\$f$ to $sender$ and terminate \;
    }
}


\Fn{ComputationComplete($idx$, $K_{result}Hash$) ~~~~ \texttt{// Callable by \mdc}}{
	$R[idx].krHash\leftarrow K_{result}Hash$ \;
	$R[idx].status\leftarrow \textsc{wait\_complete}$ \;
}

\Fn{CompleteTransaction($idx$, $K_{result}$) ~~~~ \texttt{// Callable by \mdo}}{
\If{$\textbf{Hash}(K_{result}) = R[idx].krHash$} 
    {   
        Send $\$f$ to $DO$ \;
        $R[idx].kr\leftarrow K_{result}$ \;
        $R[idx].status\leftarrow \textsc{complete}$ ~~~~\texttt{// Data transaction complete} \;
    }
}

\Fn{Cancel($idx$) ~~~~ \texttt{// Callable by \mdc}}{
	\If{$sender=R[idx].DC$ \textbf{and} $(sys.time-R[idx].reqTime) > pRTO$}
	{
	    Return $\$f$ to $R[idx].DC$; \;
	    $R[idx].status=\textsc{canceled}$ \;
	}
}


\Fn{Revoke() ~~~~ \texttt{// Callable by \mdo}}{ 
	\If{$sender=DO$}
	{ contract selfdestruct \; }
}
		
\end{algorithm}


\textbf{Data Owner's Smart Contract \ctdo~}
We design \ctdo to capture the functionalities discussed above. The pseudo code of \ctdo in shown in Algorithm~\ref{alg.dataContract}. 
In addition to the policy variables, \ctdo encodes functions
for enforcing the control logic.
\emph{Constructor} initializes the policy at contract creation. \emph{Request} takes a payment deposit from \mdc along with the requested operation $op$, the requested data descriptor $D_{target}$,
and authorizes this data transaction. \emph{ComputationComplete} is called by \mdc to signal the completion of the off-chain data execution. \emph{CompleteTransaction} is called by \mdo to record the data usage and completes the transaction. The deposited payment is then redistributed to \mdo. We will cover more details on them along with the result commitment process in Section \ref{sec:offchainexe}. \emph{Cancel} is called by \mdc to abort the current transaction if the timeout passes. Lastly, \emph{Revoke} invalidates the contract and can be called only by \mdo.

%

\subsection{Using Data Broker to Address the On-Chain Scalability Challenge}
\label{subsec:blockchainScalability}

While \ctdo allows individual \mdos to have fine-grained control over data usage policy and participate in data market independently,
this paradigm puts heavy pressure on the blockchain transaction processing capability when the number of \mdos is huge. In the meantime limited transaction throughput is a known problem for major public blockchains \cite{bitcoinRate,croman2016scaling,eyal2016bitcoin}. 
%
%
%
While there are many ongoing efforts to scale up transaction throughput \cite{poon2015bitcoin,raidenNet}, we take a different but complementary approach to address this issue in \sysname's scenario. A trusted delegate, namely data broker (\mdb), is used to represent a group of users and curates a \mdb's contract (\ctdb). 
\ctdb allows individual \mdos to register data entries and operations for \mdb to moderate. \mdb then participates in the data market on behalf of the registered \mdos. We call this paradigm the \mdb-based system in our later implementation, in contrast to the \mida-based system.

The pseudo code of \ctdb is provided in Appendix \ref{app:dbcontract}. \ctdb emulates \ctdo for most parts but with extra global variables for data source management and two more functions: \emph{Register} and \emph{Confirm}.
When a \mdo wants to make use of the \mdb, she first invokes \textit{Register} function to register her data with the \ctdb. In the data plane, the \mdo needs to remotely attest the \mdb to establish trust, then provisions the data keys to the \mdb enclave. This, however, is not the end of data registration, because the data source and quality still need to be verified by the \mdb. Once verified, \mdb invokes \textit{Confirm} function to complete the data registration. Furthermore, result commitment is also slightly different for \ctdb. The \emph{CompleteTransaction} function is now callable by \mdb and needs to distribute payments to all involved \mdos. 
%



\section{Off-Chain Contract Execution}
\label{sec:offchainexe}

%
%
%
%
%

\sysname leverages blockchain smart contract to provide the control mechanisms for valued data exchanges. While the technology offers a distributed time-stamped ledger which is ideal in providing a transparent recording of data usage, smart contract suffers from several prohibiting drawbacks when it comes to confidential data computation purely on-chain. First, the smart contract invocation and the ensuing computation is executed and repeated by all nodes in the blockchain network. The cost to run complex algorithms on-chain can be prohibitive even assuming data storage is not an issue. Second, data has to be decrypted and stored on the chain, causing confidentiality problems.

\begin{figure}
\vspace{-.2in}
\centering
 \includegraphics[width=4.7in]{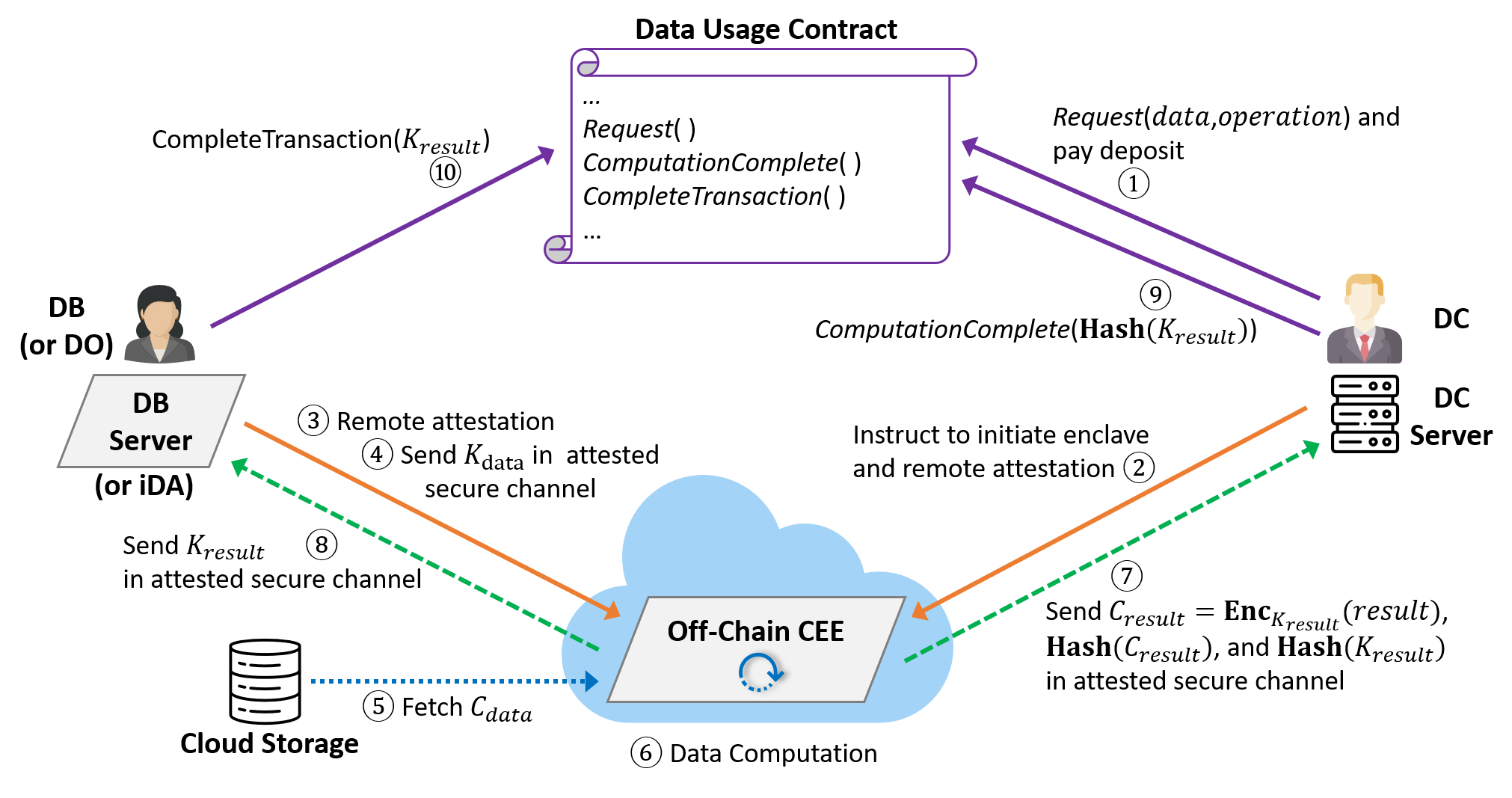}
\vspace{-0.15in}
 \caption{Off-chain Contract Execution and Result Commitment}
 \label{fig.offchainexe}
\centering
\vspace{-0.15in}
\end{figure}

To tackle this problem, we introduce the concept of \emph{off-chain contract execution} in \sysname and introduce an entity called off-chain contract execution environment (\tcee) to bring both the computation and data provisioning off-chain. Particularly, we decompose a data usage contract into two portions, the control part and the computation part. 
The control flow starts with invoking the contract and stops at the contracted computation task which switched to off-chain. The control flow is resumed with another contract invocation when the off-chain computation task is finished. 
%
Accordingly, we propose a novel off-chain contract execution and result commitment protocol, as is shown in Fig.~\ref{fig.offchainexe}. 
%
Note that both \mdb and \mida can represent a \mdo. Here we resort to the \mdb-based paradigm for convenience of presentation. We defer the discussion on \mdb's role in the data plane to the end of this section.
Next we elaborate on the important features of off-chain contract execution in a constructive manner. 

\subsection{Establishing Trust on the Execution of Contracted Operation through ``Local Consensus"}

The first challenge is the correct execution of the contracted task. 
As we have mentioned, when smart contracts are executed on blockchain platform, the correctness of the execution is guaranteed by the entire network through global consensus, which suffers from high on-chain cost.
%
Our observation is that the correctness of one particular computation instance only matters to the stakeholders of the data transaction, i.e. the \mdos, \mdb, and \mdc.
And we do not need the entire network to verify the correctness.
 
In the conventional setting of distributed computing, both the \mdc and \mdo would perform the data computation task and expect the same result from each other. However, it contradicts \mdo's goal of fine-grained control on data usage if the data are directly provided to \mdc. Instead, we rely on software remote attestation, which is a widely available primitive with TEEs~\cite{sgx,trustzone}, for securely delegating the computation task to \tcee. In this paper we opt for Intel SGX \cite{sgx}. First of all, the designated computation program should pre-ratified with its program (binary) hash published in the data contract along side ``authorized operations''. When instructed by \mdc for a specific computation task, \tcee loads the corresponding enclave program for that task.
Then the two transacting sides in the data plane, \mdb and \mdc, remotely attest the enclave program to verify its authenticity and integrity with the program hash in the contract. As a result, as shown in Fig.~\ref{fig.offchainexe}, the immediate steps after data transaction request is to have \tcee load the enclave program and \mdc and \mdb remotely attest the \tcee enclave. Once correctly \tcee enclave is verified with attestation reports, both sides of the contract can then extend their trust to \tcee, knowing the attested program will execute securely in the enclave till termination, and the computation result will be genuine even if an adversary compromises \tcee's untrusted platform (i.e., ``normal world'' in TEE terminology, which includes the operating system and non-enclave programs). And finally the result produced by \tcee will be the ``local consensus'' between the two sides.

\subsection{Enforcing Data Obligation and Confidentiality}

The local consensus mechanism guarantees the data intensive computation task can be offloaded to the off-chain entity \tcee for execution while maintaining the correctness of computation. However, in order to achieve the privacy goals of \sysname, computation itself has to fulfill the \emph{data obligation}, which we refer to as the obligations of \mdc for utilizing \mdo's data. More specifically, it follows the general requirement of secure computation, wherein only the computation result is accessible by the \mdc, not the plaintext source data. 
First, the computation process should not output any plaintext source data or any intermediate results that are derived from the source data. Second, at the end of the computation, all decrypted data and intermediate results should be sanitized.
Despite recent breakthrough in fully homomorphic encryption, performing arbitrary computation over encrypted data remains impractical for generic computation. In \sysname, we make use of TEE enclaves to create the environment for confidential computing. As is illustrated by step 3 and 4 in Fig. \ref{fig.offchainexe}, \mdo's data en/decryption key $K_{data}$ can be provisioned to \tcee's enclave only if the latter can be cryptographically verified via remote attestation and a secure channel is established. This comes as an integral part of the local consensus. The hardware of \tcee, the processor specifically, enforces the isolation between the untrusted platform and the enclave. We require the enclave program to include steps to sanitize intermediate results and keying materials. Since memory contents are encrypted in Intel SGX, once the keying material is removed, the data can be considered effectively sanitized. This also ensures that the program inside the enclave will terminate once the contracted task is completed. 



\subsection{Ensuring Atomicity in Contract Execution and Result Commitment}

The last challenge is ensuring the atomicity of the contract, which arises from the split of control between on-chain off-chain. Contract functions that were previously executed in a single block are now completed via multiple function invocations that are executed in multiple blocks. Furthermore, there is no guarantee on the execution time of the off-chain computation, because an adversary controlling the platform can interrupt the computation and cause delays. 
Specifically, two issues need to be addressed.

The first issue is the contract function runtime. When the adversary has control of the off-chain computation platform of \tcee, he can pause or delay the computation. For many data computations, the result can be time-sensitive. 
To tackle this problem, we add a timeout mechanism in the data contract to allow \mdc to cancel the request after timeout and have the deposit refunded (see Algorithm \ref{alg.dataContract}). 

The second issue is the atomic completion of the contract. We want both the \mdos to get the payment in the control plane while allowing \mdc to get the computation results in the data plane. This is particularly challenging due to the lack of availability guarantee on the \tcee platform. 
When the platform is compromised, the adversary can intercept and modify any external I/O from the enclave, including both the network and storage. 
Our design for the atomic completion and result commitment can be observed from step 7 to 10 in Fig.~\ref{fig.offchainexe}. The key idea is that result release and contract completion should be done as a single message in the control plane. To prevent \mdc from getting the result without completing the payment to \mdos, the result are encrypted into $C_{result}$ with a random result key $K_{result}$, before being sent to \mdc in the attested secure channel. Since the platform can corrupt any output from \tcee, the \tcee enclave also sends \mdc the hash of the encrypted result and key, i.e.,  $\mathbf{Hash}(C_{result})$ and $\mathbf{Hash}(K_{result})$, which will be later used by \mdc for integrity check on the result and the key. $K_{result}$ is passed to \mdb in the attested secure channel. 
To prevent \mdb from completing the transaction without releasing the correct result key,
\mdc needs to initiate the commitment procedure in the control plane by
invoking the contract function \emph{ComputationComplete} with $\mathbf{Hash}(K_{result})$, indicating it has the encrypted result and is ready to finish the data transaction if and only if the correct result key $K_{result}$ is released. Upon observing the message from \mdc, \mdb then invokes the smart contract function \emph{CompleteTransaction} with the result key $K_{result}$. Only when the hash of $K_{result}$ matches the previously received $\mathbf{Hash}(K_{result})$, will the contract write the data usage into records, release the payment to \mdos, and finally conclude the data transaction. 
Note that our commitment protocol design does not need to protect the confidentiality of $K_{result}$ (thus enabling the on-chain hash check). This is because the encrypted result $C_{result}$ is passed directly from \tcee enclave to \mdc via the attested secure channel. Finally, \mdc has the full discretion in deciding whether to publish the computation result afterwards.

\subsection{Data Broker for Scalability in the Data Plane}
\label{subsec:db-scale-data-plane}

In the \mida-based paradigm, when \mdc needs to use the data from a large number of \mdos, the naive use of remote attestation on the \tcee would require each \mida to individually attest and verify the \tcee enclave, resulting in linearly growing computation overhead and network traffic. To address this challenge, in the \mdb-based paradigm, \mdb can be re-purposed as a trusted intermediary between the \tcee and all relevant \mdos in the data plane during the preparation stage, similar to its control plane role.
Essentially, \mdb is also deployed on a TEE-enabled machine and instantiates an enclave for secure handling of \mdos' data. The enclave is attested to every new \mdo only once after the \mdo registers with \mdb. During the normal operation, \mdb attests \tcee on behalf of all relevant \mdos for each \mdc request, saving the need for individual \mdos to attest \tcee.
To accommodate the extreme case when a large number of \mdos registers with \mdb simultaneously, we will explore parallel remote attestation solutions in Subsection \ref{subsec:control-plane-runtimes}.
\section{Implementation and Evaluation}












We implemented a prototype of \sysname using Intel SGX as the TEE technology and Ethereum as the smart contract platform. Source code with documentation is available at \url{https://github.com/yang-sec/PrivacyGuard}.
The on-chain components, namely the \mdo contract and the \mdb contract, were implemented in Solidity with 144 and 162 software lines of code (SLOC) respectively. The data usage price was set at $0.01$ ethers per user data.
The off-chain components include five \sysname applications, namely \emph{\ida (iDA)}, \emph{Data Broker (DB)}, \emph{Data Owner (DO)}, \emph{Data Consumer (DC)}, \emph{Contract Execution Environment (CEE)}. They were implemented in C++ with Intel SGX SDK v2.3.1 on top of Ubuntu 16.04 LTS. The total SLOC for off-chain components is about 37,000.
We deployed the contracts onto Ethereum Rinkeby testnet for evaluation, though our system is fully compatible with Ethereum mainnet. We used a fixed gas price of $10^{-9}$ ethers.
PrivacyGuard applications were deployed in a LAN scenario. 
1 \mdb, 1 \ida, and 1 \tcee ran on a SGX-enabled Linux machine with Intel Core i5-7260U CPU (2 cores 4 threads, 3.5 GHz).
Up to $160$ \mdos and 1 \mdc ran on a Linux machine with AMD FX-8320 CPU (4 cores 8 threads, 3.5 GHz).
%
We note that this setup aims for feasibility demonstration; in real-world deployment each application will most likely reside in a different machine.
%
%
%
We used the \textit{adult} dataset from UCI Machine Learning Repository \cite{Dua:2017}
to simulate the data source.
Each \mdo randomly drew 500 data points from the dataset as its private data.
%
We have tested the entire PrivacyGuard workflow in multiple runs and the data usage history has been recorded in the deployed contracts.
Our evaluation focuses on the system's scalability and consists of three parts: control plane runtimes, control plane costs, and data plane runtimes.

\subsection{Control Plane Runtimes}
\label{subsec:control-plane-runtimes}

\begin{figure}
    \vspace{-.15in}
    \centering
    \subfigure[]{
        \centering
        \includegraphics[width = 2.3in]{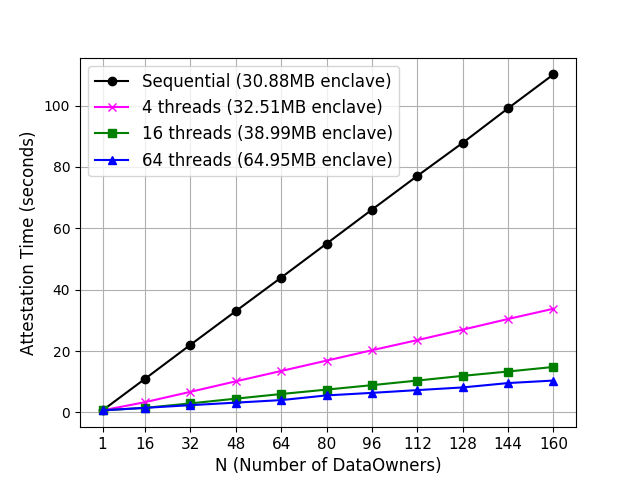}
        \label{fig:attestation-times}
    }
    \subfigure[]{
        \centering
        \includegraphics[width = 2.3in]{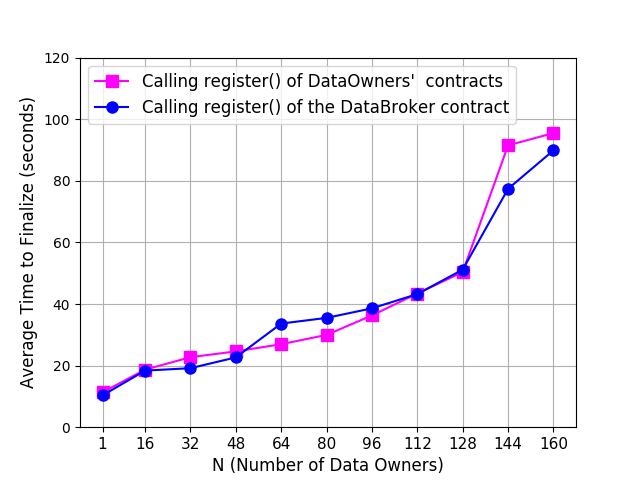}
        \label{fig:contract-call-times}
    }
\vspace{-0.1in}
 \caption{
 \footnotesize
 (a) Attestation times of \mdb when $N$ \mdos simultaneously initiate attestation. 
 (b) Average transaction finalization delay when $N$ \mdos simultaneously call a contract function.}
 \vspace{-0.1in}
\centering
\end{figure}

To accommodate the scenario where $N$ \mdos simultaneously attest the \mdb enclave in the \mdb-based system, we experimented with a parallel attestation scheme in \mdb that each of the $N$ attestation instances is handled by one of the $T$ software threads, which invokes a new attestation context of the enclave and a dedicated enclave thread control structure (TCS) (thus TCSNUM = $T$). The experiment was repeated under different $T$. To avoid congesting the Intel Attestation Service (IAS) which may violate the terms of service, we instead used a simulated IAS that responds to EPID signature revocation list request and attestation report request with 0.1s and 0.5s delays respectively. The result is shown in Fig. \ref{fig:attestation-times}. We observe that the parallel scheme is indeed a promising solution for scaling up attestation capacity, at the cost of enlarged enclave memory footprint. When $N=160$, it takes the 64-thread \mdb about a tenth the attestation time of its sequential counterpart. We remark that efficient and scalable remote attestation is an interesting standalone topic to explore in future work.

To further evaluate the performance constraints imposed by the blockchain network, we measured the average transaction finalization delay in a congested environment. We set up 160 \mdos to simultaneously send out a transaction calling the \emph{Register()} function in the \mdb contract and their own \mdo contracts.
we use \emph{receipt} as the finalization response of the Ethereum transaction that makes the function call.
The result is shown in Fig. \ref{fig:contract-call-times}.
As more \mdos send transactions at the same time, the average time to finalize a transaction increases dramatically. A straightforward workaround is to require \mdos to call \emph{Register()} according to a time schedule that minimizes congestion.



\subsection{Control Plane Cost}










The monetary cost of the control plane mainly comes from the gas cost of operating smart contracts in Ethereum. At the beginning, every \mdo registers its data items on its own \mdo contract and the \mdb contract. \mdb fetches data from whoever registered with its contract and routinely confirms new registries. \mdc then requests for the data items from $N$ \mdos by sending a request transaction to the \mdb contract (or separate requests to all related \mdo contracts) with a sufficient deposit to cover the price before proceeding to attesting \tcee. We repeated the experiment for $N=1\rightarrow10$ and obtained the gas costs and dollar equivalents for each contract function call, based on the ether price on 03/31/2019, which was \$141.51 (source: \url{https://coinmarketcap.com/}).


\begin{table}[h]
    \vspace{-.1in}
    \centering
    \small
    \caption{Cost of the data contract's scale-independent functions}
    \begin{tabular}{l|rc|rc}
        \toprule
        & \multicolumn{2}{c|}{\mdo Contract} & \multicolumn{2}{c}{\mdb Contract} \\
        Function  & Gas Cost  & USD Equiv. & Gas Cost  & USD Equiv. \\ \hline
        constructor()  & $951747$ & $0.13468$ & $846794$ & $0.11983$ \\
        Register() (new)  & $156414$ & $0.02213$ & $125392$ & $0.01774$ \\
        Register() (update)  & $30121$ & $0.00426$ & $45177$ & $0.00639$ \\
        Cancel() & $81998$ & $0.01160$ & $66954$ & $0.00947$ \\
        \bottomrule
    \end{tabular}
    \vspace{-0in}
    \label{tab:cost-scale-free-calls}
\end{table}

\begin{figure}
    \vspace{-.2in}
    \centering
        \subfigure[]{
            \centering
            \includegraphics[width = 2.3in]{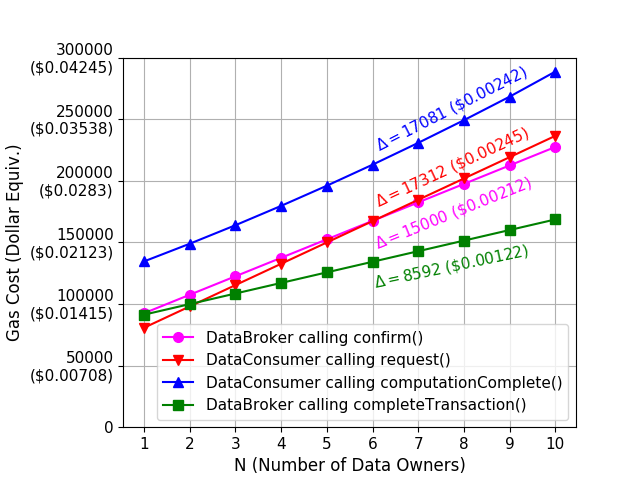}
            \label{fig:cost-contract-func-calls-a}
        }
        \subfigure[]{
            \centering
            \includegraphics[width = 2.3in]{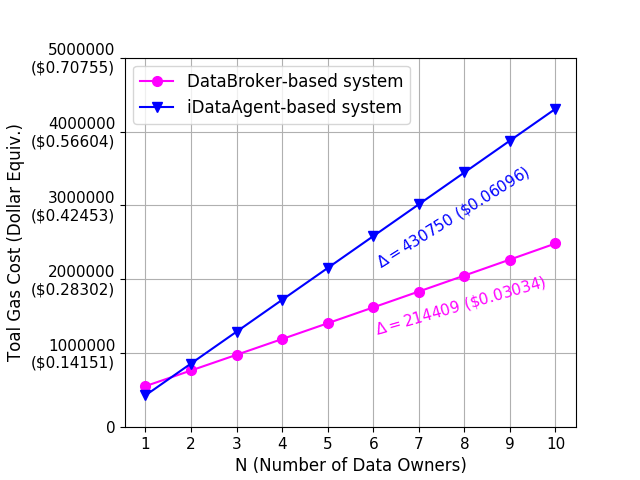}
            \label{fig:cost-contract-func-calls-b}
        }
    \vspace{-0.1in}    
    \caption{
    \footnotesize
    (a) Gas costs of the \mdb contract's scale-dependent function calls. (b) Total gas costs of the \mdb based system and the \ida based system. 
    }
    \label{fig:cost-contract-func-calls}
    \vspace{-0.20in}
\end{figure}

We find that in both \mdb and \mdo contracts the costs of calling \emph{constructor()} (contract creation), \textit{Register()} and \textit{Cancel()} do not depend on the number of registered \mdos. We call these type of function calls scale-independent; otherwise scale-dependent.
The costs of calling scale-independent functions and scale-dependent functions are shown in Table \ref{tab:cost-scale-free-calls} and Fig. \ref{fig:cost-contract-func-calls-a}. 
Notably, the costs of calling \textit{Request()} and \textit{ComputationComplete()} grow faster than the costs of calling \textit{Confirm()} and \textit{CompleteTransaction()}. This implies the total cost will increasingly shift to the \mdc side, which is a scalable trend for the system as the \mdc has incentives to pay for more data usage.

To evaluate the scalability gain brought by \mdb, we compare the case wherein individual \mdos share data via their own \mdo contracts versus via the \mdb contract. In both cases, the total amount of data requested by the \mdc and subsequently operated with by the \tcee are the same. 
We summed the costs of all function calls except for the contract creation (calling \textit{constructor()}) and extrapolated over different $N$. Fig. \ref{fig:cost-contract-func-calls-b} shows that it costs the \mdb-based system much less to accommodate one extra \mdo ($\$0.0304$) compared to the \ida-based system ($\$0.06096$). 
This result together with control plane runtimes (Fig. \ref{fig:attestation-times}) demonstrate \mdb's ability to provide \sysname with financial and performance scalability when facing a growing number of \mdos.



\subsection{Data Plane Runtimes}
\label{subsec:data-plane-runtimes}



To evaluate \tcee's performance in off-chain contract execution,
we experimented with a demonstrative, reasonably complex computation task: training four parallel instances of a neural network classifier. Detailed hyperparameters can be found in our source code. The training functions were ported to the SGX enclave from the Fast Artificial Neural Network (FANN) Library (\url{https://github.com/libfann/fann}). 
%
To evaluate enclave overhead, we also implemented an untrusted version (executed outside enclave) of the computation task that ran on the same machine. 
We noticed that recent work showed Intel's Hyperthreading Technology (HTT) has flaws that may impair the security of SGX enclaves \cite{van18foreshadow}.
%
Therefore, we tested the computation task under different hardware options with respect to the usage of SGX enclave and HTT. The Intel CPU's TurboBoost feature was turned off to avoid unexpected performance gain.

\begin{figure}
    \vspace{-.2in}
    \centering
    \includegraphics[width = 2.7in]{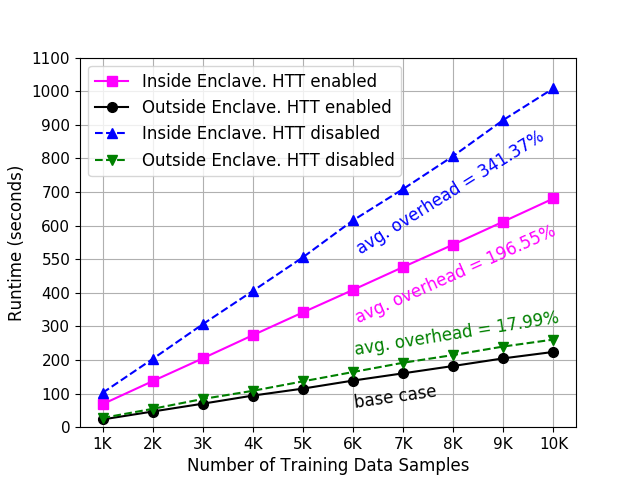}
    \caption{Runtimes of training an example neural network classifier under four hardware options.}
    \label{fig:data-plane-computation-time}
\end{figure}

The experiment results is shown in Fig. \ref{fig:data-plane-computation-time}.
We find that the overhead caused by disabling HTT is $48.84\%$ for inside enclave and  $17.99\%$ for outside-enclave. This indicates disabling HTT will drag down in-enclave performance more significantly. The overheads caused by enclave are $196.55\%$ and $274.13\%$ for HTT-enabled and HTT-disabled respectively. 
We speculate that the big enclave overhead is related to the enclave's secure function calls and our imperfect porting of the training program.
We leave the performance caveats of Intel SGX and possible solutions to future work.

\section{Related Work}
\label{sec.relatedWork}


\textbf{Privacy Protection}
Privacy-preserving computation has been an active area of research in the past decade~\cite{ryoan,vc3,ohrimenko2016oblivious,cao2014privacy,verykios2004state,zhang2018privacyguard}. With the increasing reliance on rich data, there has been a significant amount of research on applying cryptographic techniques to perform privacy preserving computation and data access control~\cite{pass2017formal,boneh2011functional,cao2014privacy,verykios2004state,vimercati2010encryption,bacis2016mix}. Recently, hardware-assisted TEE has been adapted in numerous works to achieve privacy-preserving computation~\cite{ohrimenko2016oblivious,ryoan,vc3,opaque,hunt2018chiron,fisch2017iron}. Specially,
Ryoan~\cite{ryoan} is closely related to \sysname. It combines native client sandbox and Intel SGX to confine data processing module and provide confidentiality. 
However, Ryoan aims to achieve data confinement with a user-defined directed acyclic graph that specifies information flow. In comparison, \sysname allows data user and consumer to negotiate data usage using smart contract with non-repudiable usage recording. 



\noindent
\textbf{Blockchain and TEE}
The idea of moving computation off-chain to improve the performance and security is mentioned in ~\cite{choudhuri2017fairness,intelPDO,ekiden,kalodner2018arbitrum,sinhaluciditee,wust2019ace}. Choudhuri et al.~\cite{choudhuri2017fairness} combines blockchain with TEE to build one-time programs that resemble to smart contracts but only aim for a restricted functionality.
%
%
Ekiden~\cite{ekiden} and the Intel Private Data Object (PDO) project~\cite{intelPDO} are two concurrently developed projects that are closely related to our work. Similar to \sysname, Ekiden harmonizes trusted computing and distributed ledger to enable confidential contract execution.
Ekiden offloads computation from consensus nodes to a collection computing nodes in the aim of improving the ecosystem.
In comparison, \sysname is designed
to fit existing blockchain infrastructure. 
The Intel PDO project aims to combine Intel SGX and distributed ledger to allow distrustful parties to work on the data in a confidential manner. However, the system focuses heavily on a permissioned model with significant overhead for bootstrapping trust.

\section{Conclusion}

In this paper, we proposed \sysname, a platform that combines blockchain smart contract and TEE to enable transparent enforcement of private data computation and fine-grained usage control.
%
Blockchain can not only be used as a tamper-proof distributed ledger that records data usage, but also facilitate financial transactions to incentivize data sharing. To enable complex and confidential operations on private data, \sysname splits smart contract functionalities into control operations and data operations. Remote attestation and TEE are used to achieve local consensus of the contract participants on the trustworthiness of the off-chain contract execution environment. Atomicity of the contract completion and result release is facilitated by a commitment protocol. 
We implemented a prototype of \sysname platform and evaluated it in a simulated data market. The results show the reasonable control plane costs and feasibility of executing complex data operations in a confidential manner using the platform.


\section*{ACKNOWLEDGMENT}
This work was supported in part by US National Science Foundation under grants CNS-1916902 and CNS-1916926.

\appendix



\section{Data Broker Contract \ctdb}
\label{app:dbcontract}

\vspace{-.3in}

\SetKwInput{KwInit}{Init}
\SetKwInput{KwRequestDataUse}{RequestDataUse}
\SetKwInput{KwFreeze}{Freeze}
\SetKwInput{KwCompute}{Compute}
\SetKwInput{KwFinalize}{Finalize}
\SetKwInput{KwCancel}{Cancel}
\SetKwInput{KwRecord}{Record}
\SetKwInput{KwRevokeContract}{RevokeContract}

\SetKwInput{KwRegisterData}{RegisterUserDataSource}
\SetKwInput{KwRegister}{Register}
\SetKwInput{KwConfirmData}{ConfirmUserDataSourceRegistry}
\SetKwInput{KwConfirm}{Confirm}
\SetKwInput{KwCompComplete}{ComputationComplete}
\SetKwInput{KwCompleteTrans}{CompleteTransaction}

\begin{algorithm}
    \algfootnote
	\caption{Data Broker's Smart Contract \ctdb Pseudocode}
	\label{alg.db.contract}
	\scriptsize


    \Fn{Constructor() ~~~~ \texttt{// Contract creation by \mdb with config}}{
		Parse $config$ as ($operationList, requestTimeout$)          \;
		$cOPL    				\leftarrow config.operationList$ 			\;
		$cRTO 				    \leftarrow config.requestTimeout$ 			\;
		$\{DO,DS,R\}  \leftarrow \{[[~]],[~],[~]\}$   ~~~~ \texttt{// \mdos, data sources, data usage records} \;
		$DB \leftarrow creator$ \;
	}
	
 	

\Fn{Register($op,DC,price$) ~~~~ \texttt{// Callable by a \mdo}}{
    Create a \mdo entry $DO[ido,op]$ with index $ido$ for this new \mdo \;
    $DO[ido,op].\{DO,DC,price\}\leftarrow \{sender,DC,price\}$ \;
}


\Fn{Confirm($cfDOs$) ~~~~ \texttt{// Callable by \mdb}}{
    \For{\textbf{all} $\{ido,op\}$ \textbf{that} $ido\in cfDOs$ \textbf{and} $op\in cOPL$ \textbf{and} $DO[ido,op]\neq null$}{
        Append $ido$ to $DS[op].DOList$ \;
        Append $DO[ido,op].DC$ to $DS[op].DCList$ \;
        $DS[op].price \leftarrow DS[op].price + DO[ido,op].price$ \;
    }
}

\Fn{Request($op,targetDOs,\$f$) ~~~~ \texttt{// Callable by \mdc}}{
    \eIf{$op\in cOPL$ \textbf{and} $sender\in DS[op].DCList$ \textbf{and} $targetDOs\subset DS[op].DOList$ \textbf{and} $f\geq DS[op].price$ }
	{  
	    Create a record entry $R[idx]$ with index $idx$ for this new data transaction \;
	    $R[idx].\{targetDOs,DC,reqTime\}\leftarrow \{targetDOs,sender,sys.time\}$ \;
	    $R[idx].status\leftarrow\textsc{wait\_computation}$ \;
	}
    {
        Return $\$f$ to $sender$ and terminate \;
    }
}





\Fn{ComputationComplete($idx,K_{result}Hash$)}{(same as in \ctdo, see Algorithm \ref{alg.dataContract})}

\Fn{CompleteTransaction($idx$, $K_{result}$) ~~~~ \texttt{// Callable by \mdb}}{
\If{$\textbf{Hash}(K_{result}) = R[idx].krHash$} 
    {   
        \For{\textbf{all} $ido\in DS[R[idx].op].DOList$}{
            Send $\$DO[ido,R[idx].op].price$ to $DO[ido].DO$\;
        }
        $R[idx].kr\leftarrow K_{result}$ \;
        $R[idx].status\leftarrow \textsc{complete}$ ~~~~\texttt{// Data transaction complete} \;
    }
}

    
 	
	\Fn{Cancel($idx$)}{(same as in \ctdo)}
	\Fn{Revoke()}{(same as in \ctdo, except callable by \mdb)}


\end{algorithm}

%
%
%
%

\bibliographystyle{splncs04}
\bibliography{references}{}

\begin{thebibliography}{10}
\providecommand{\url}[1]{\texttt{#1}}
\providecommand{\urlprefix}{URL }
\providecommand{\doi}[1]{https://doi.org/#1}

\bibitem{raidenNet}
brainbot~technologies AG: Raiden network. \url{https://https://raiden.network/}

\bibitem{trustzone}
{ARM}: Security technology building a secure system using trustzone technology
  (2009)

\bibitem{bacis2016mix}
Bacis, E., De~Capitani~di Vimercati, S., Foresti, S., Paraboschi, S., Rosa, M.,
  Samarati, P.: Mix\&slice: Efficient access revocation in the cloud. In:
  Proceedings of the 2016 ACM SIGSAC Conference on Computer and Communications
  Security. pp. 217--228 (2016)

\bibitem{bethencourt2007ciphertext}
Bethencourt, J., Sahai, A., Waters, B.: Ciphertext-policy attribute-based
  encryption. In: 2007 IEEE symposium on security and privacy (SP'07). pp.
  321--334. IEEE (2007)

\bibitem{boneh2011functional}
Boneh, D., Sahai, A., Waters, B.: Functional encryption: Definitions and
  challenges. In: Theory of Cryptography Conference. pp. 253--273. Springer
  (2011)

\bibitem{intelPDO}
Bowman, M., Miele, A., Steiner, M., Vavala, B.: Private data objects: an
  overview (2018), \url{https://arxiv.org/pdf/1807.05686.pdf}

\bibitem{bitcoinRate}
Buterin, V.: Privacy on blockchain.
  \url{https://blog.ethereum.org/2016/01/15/privacy-on-the-blockchain/}

\bibitem{cao2014privacy}
Cao, N., Wang, C., Li, M., Ren, K., Lou, W.: Privacy-preserving multi-keyword
  ranked search over encrypted cloud data. IEEE Transactions on parallel and
  distributed systems  \textbf{25}(1),  222--233 (2014)

\bibitem{ekiden}
Cheng, R., Zhang, F., Kos, J., He, W., Hynes, N., Johnson, N., Juels, A.,
  Miller, A., Song, D.: Ekiden: A platform for confidentiality-preserving,
  trustworthy, and performant smart contracts. In: 2019 IEEE European Symposium
  on Security and Privacy (EuroS\&P). pp. 185--200. IEEE (2019)

\bibitem{choudhuri2017fairness}
Choudhuri, A.R., Green, M., Jain, A., Kaptchuk, G., Miers, I.: Fairness in an
  unfair world: Fair multiparty computation from public bulletin boards. In:
  Proceedings of the 2017 ACM SIGSAC Conference on Computer and Communications
  Security. pp. 719--728. ACM (2017)

\bibitem{costan2016sanctum}
Costan, V., Lebedev, I.A., Devadas, S.: Sanctum: Minimal hardware extensions
  for strong software isolation. In: USENIX Security Symposium. pp. 857--874
  (2016)

\bibitem{croman2016scaling}
Croman, K., Decker, C., Eyal, I., Gencer, A.E., Juels, A., Kosba, A., Miller,
  A., Saxena, P., Shi, E., Sirer, E.G., et~al.: On scaling decentralized
  blockchains. In: International Conference on Financial Cryptography and Data
  Security. pp. 106--125. Springer (2016)

\bibitem{datareuseTaxonomyYaleLaw}
Custers, B., Ur{\v{s}}i{\v{c}}, H.: Big data and data reuse: a taxonomy of data
  reuse for balancing big data benefits and personal data protection.
  International Data Privacy Law  \textbf{6}(1),  4--15 (2016)

\bibitem{datta2017use}
Datta, A., Fredrikson, M., Ko, G., Mardziel, P., Sen, S.: Use privacy in
  data-driven systems: Theory and experiments with machine learnt programs. In:
  Proceedings of the 2017 ACM SIGSAC Conference on Computer and Communications
  Security. pp. 1193--1210. ACM (2017)

\bibitem{Dua:2017}
Dheeru, D., Karra~Taniskidou, E.: {UCI} machine learning repository (2017),
  \url{http://archive.ics.uci.edu/ml}

\bibitem{dwork2008differential}
Dwork, C.: Differential privacy: A survey of results. In: International
  conference on theory and applications of models of computation. pp. 1--19.
  Springer (2008)

\bibitem{elnikety2016thoth}
Elnikety, E., Mehta, A., Vahldiek-Oberwagner, A., Garg, D., Druschel, P.:
  Thoth: Comprehensive policy compliance in data retrieval systems. In: USENIX
  Security Symposium. pp. 637--654 (2016)

\bibitem{EthereumWeb}
Ethereum: Blockchain app platform. \url{https://www.ethereum.org/}

\bibitem{eu2016gdpr}
General data protection regulation ({GDPR}) (2016),
  \url{https://eur-lex.europa.eu/eli/reg/2016/679/oj}

\bibitem{eyal2016bitcoin}
Eyal, I., Gencer, A.E., Sirer, E.G., Van~Renesse, R.: {Bitcoin-NG}: A scalable
  blockchain protocol. In: NSDI. pp. 45--59 (2016)

\bibitem{facebookScandalWeb}
Facebook–cambridge analytica data scandal.
  \url{https://en.wikipedia.org/wiki/Facebook%E2%80%93Cambridge_Analytica_data_scandal}

\bibitem{fisch2017iron}
Fisch, B., Vinayagamurthy, D., Boneh, D., Gorbunov, S.: Iron: functional
  encryption using intel sgx. In: Proceedings of the 2017 ACM SIGSAC Conference
  on Computer and Communications Security. pp. 765--782. ACM (2017)

\bibitem{goyal2006attribute}
Goyal, V., Pandey, O., Sahai, A., Waters, B.: Attribute-based encryption for
  fine-grained access control of encrypted data. In: Proceedings of the 13th
  ACM conference on Computer and communications security. pp. 89--98 (2006)

\bibitem{hunt2018chiron}
Hunt, T., Song, C., Shokri, R., Shmatikov, V., Witchel, E.: Chiron:
  Privacy-preserving machine learning as a service (2018),
  \url{https://arxiv.org/pdf/1803.05961.pdf}

\bibitem{ryoan}
Hunt, T., Zhu, Z., Xu, Y., Peter, S., Witchel, E.: Ryoan: A distributed sandbox
  for untrusted computation on secret data. In: OSDI. pp. 533--549 (2016)

\bibitem{kalodner2018arbitrum}
Kalodner, H., Goldfeder, S., Chen, X., Weinberg, S.M., Felten, E.W.: Arbitrum:
  Scalable, private smart contracts. In: Proceedings of the 27th USENIX
  Conference on Security Symposium. pp. 1353--1370. USENIX Association (2018)

\bibitem{hawk}
Kosba, A., Miller, A., Shi, E., Wen, Z., Papamanthou, C.: Hawk: The blockchain
  model of cryptography and privacy-preserving smart contracts. In: Security
  and Privacy (SP), 2016 IEEE Symposium on. pp. 839--858. IEEE (2016)

\bibitem{li2007t}
Li, N., Li, T., Venkatasubramanian, S.: t-closeness: Privacy beyond k-anonymity
  and l-diversity. In: 2007 IEEE 23rd International Conference on Data
  Engineering. pp. 106--115. IEEE (2007)

\bibitem{machanavajjhala2007diversity}
Machanavajjhala, A., Kifer, D., Gehrke, J., Venkitasubramaniam, M.:
  l-diversity: Privacy beyond k-anonymity. ACM Transactions on Knowledge
  Discovery from Data (TKDD)  \textbf{1}(1),  3--es (2007)

\bibitem{sgx}
McKeen, F., Alexandrovich, I., Berenzon, A., Rozas, C.V., Shafi, H.,
  Shanbhogue, V., Savagaonkar, U.R.: Innovative instructions and software model
  for isolated execution. In: HASP@ ISCA. p.~10 (2013)

\bibitem{nakamoto2008bitcoin}
Nakamoto, S.: Bitcoin: A peer-to-peer electronic cash system (2008)

\bibitem{NPRS2016}
{National Science and Technology Council}: National privacy research strategy,
  \url{https://www.nitrd.gov/PUBS/NationalPrivacyResearchStrategy.pdf}

\bibitem{nissenbaum2004privacy}
Nissenbaum, H.: Privacy as contextual integrity. Washington Law Review
  \textbf{79}, ~119 (2004)

\bibitem{ohrimenko2016oblivious}
Ohrimenko, O., Schuster, F., Fournet, C., Mehta, A., Nowozin, S., Vaswani, K.,
  Costa, M.: Oblivious multi-party machine learning on trusted processors. In:
  USENIX Security Symposium. pp. 619--636 (2016)

\bibitem{pass2017formal}
Pass, R., Shi, E., Tramer, F.: Formal abstractions for attested execution
  secure processors. In: Annual International Conference on the Theory and
  Applications of Cryptographic Techniques. pp. 260--289. Springer (2017)

\bibitem{poon2015bitcoin}
Poon, J., Dryja, T.: The bitcoin lightning network: Scalable off-chain instant
  payments (2016),
  \url{https://www.bitcoinlightning.com/wp-content/uploads/2018/03/lightning-network-paper.pdf}

\bibitem{vc3}
Schuster, F., Costa, M., Fournet, C., Gkantsidis, C., Peinado, M., Mainar-Ruiz,
  G., Russinovich, M.: Vc3: Trustworthy data analytics in the cloud using sgx.
  In: Security and Privacy (SP), 2015 IEEE Symposium on. pp. 38--54. IEEE
  (2015)

\bibitem{sen2014bootstrapping}
Sen, S., Guha, S., Datta, A., Rajamani, S.K., Tsai, J., Wing, J.M.:
  Bootstrapping privacy compliance in big data systems. In: Security and
  Privacy (SP), 2014 IEEE Symposium on. pp. 327--342. IEEE (2014)

\bibitem{sinhaluciditee}
Sinha, R., Gaddam, S., Kumaresan, R.: Luciditee: Policy-compliant fair
  computing at scale (2019), \url{https://eprint.iacr.org/2019/178.pdf}

\bibitem{song2019sok}
Song, D., Lettner, J., Rajasekaran, P., Na, Y., Volckaert, S., Larsen, P.,
  Franz, M.: Sok: sanitizing for security. In: 2019 IEEE Symposium on Security
  and Privacy (SP). pp. 1275--1295. IEEE (2019)

\bibitem{sweeney2002k}
Sweeney, L.: k-anonymity: A model for protecting privacy. International Journal
  of Uncertainty, Fuzziness and Knowledge-Based Systems  \textbf{10}(05),
  557--570 (2002)

\bibitem{szabo1997formalizing}
Szabo, N.: Formalizing and securing relationships on public networks. First
  Monday  \textbf{2}(9) (1997)

\bibitem{tedTalkCayla}
{TED Talk}: How tech companies deceive you into giving up your data and
  privacy. \url{https://goo.gl/hSfaUX}

\bibitem{timcookTalk}
Tim cook: Personal data collection is being `weaponized against us with
  military efficiency'. \url{https://goo.gl/BsWB3k}

\bibitem{van18foreshadow}
Van~Bulck, J., Piessens, F., Strackx, R.: Foreshadow: Extracting the keys to
  the intel $\{$SGX$\}$ kingdom with transient out-of-order execution. In: 27th
  {USENIX} Security Symposium ({USENIX} Security 18) (2018)

\bibitem{van2017telling}
Van~Bulck, J., Weichbrodt, N., Kapitza, R., Piessens, F., Strackx, R.: Telling
  your secrets without page faults: Stealthy page table-based attacks on
  enclaved execution. In: 26th {USENIX} Security Symposium ({USENIX} Security
  17). pp. 1041--1056 (2017)

\bibitem{verykios2004state}
Verykios, V.S., Bertino, E., Fovino, I.N., Provenza, L.P., Saygin, Y.,
  Theodoridis, Y.: State-of-the-art in privacy preserving data mining. ACM
  Sigmod Record  \textbf{33}(1),  50--57 (2004)

\bibitem{vimercati2010encryption}
Vimercati, S.D.C.D., Foresti, S., Jajodia, S., Paraboschi, S., Samarati, P.:
  Encryption policies for regulating access to outsourced data. ACM
  Transactions on Database Systems (TODS)  \textbf{35}(2), ~12 (2010)

\bibitem{wang2010hierarchical}
Wang, G., Liu, Q., Wu, J.: Hierarchical attribute-based encryption for
  fine-grained access control in cloud storage services. In: Proceedings of the
  17th ACM conference on Computer and communications security. pp. 735--737
  (2010)

\bibitem{wang2017leaky}
Wang, W., Chen, G., Pan, X., Zhang, Y., Wang, X., Bindschaedler, V., Tang, H.,
  Gunter, C.A.: Leaky cauldron on the dark land: Understanding memory
  side-channel hazards in sgx. In: Proceedings of the 2017 ACM SIGSAC
  Conference on Computer and Communications Security. pp. 2421--2434. ACM
  (2017)

\bibitem{wust2019ace}
W{\"u}st, K., Matetic, S., Egli, S., Kostiainen, K., Capkun, S.: Ace:
  Asynchronous and concurrent execution of complex smart contracts. (2019),
  \url{https://eprint.iacr.org/2019/835.pdf}

\bibitem{xu2015controlled}
Xu, Y., Cui, W., Peinado, M.: Controlled-channel attacks: Deterministic side
  channels for untrusted operating systems. In: 2015 IEEE Symposium on Security
  and Privacy. pp. 640--656. IEEE (2015)

\bibitem{yu2010achieving}
Yu, S., Wang, C., Ren, K., Lou, W.: Achieving secure, scalable, and
  fine-grained data access control in cloud computing. In: Infocom, 2010
  proceedings IEEE. pp.~1--9. Ieee (2010)

\bibitem{zhang2016town}
Zhang, F., Cecchetti, E., Croman, K., Juels, A., Shi, E.: Town crier: An
  authenticated data feed for smart contracts. In: Proceedings of the 2016 ACM
  SIGSAC Conference on Computer and Communications Security. pp. 270--282. ACM
  (2016)

\bibitem{zhang2018privacyguard}
Zhang, N., Li, J., Lou, W., Hou, Y.T.: Privacyguard: Enforcing private data
  usage with blockchain and attested execution. In: Data Privacy Management,
  Cryptocurrencies and Blockchain Technology, pp. 345--353. Springer (2018)

\bibitem{zhang2016cachekit}
Zhang, N., Sun, H., Sun, K., Lou, W., Hou, Y.T.: Cachekit: Evading memory
  introspection using cache incoherence. In: 2016 IEEE European Symposium on
  Security and Privacy (EuroS\&P). pp. 337--352. IEEE (2016)

\bibitem{zhang2016case}
Zhang, N., Sun, K., Lou, W., Hou, Y.T.: Case: Cache-assisted secure execution
  on arm processors. In: 2016 IEEE Symposium on Security and Privacy (SP). pp.
  72--90. IEEE (2016)

\bibitem{zhang2018trusense}
Zhang, N., Sun, K., Shands, D., Lou, W., Hou, Y.T.: Trusense: Information
  leakage from trustzone. In: IEEE INFOCOM 2018-IEEE Conference on Computer
  Communications. pp. 1097--1105. IEEE (2018)

\bibitem{opaque}
Zheng, W., Dave, A., Beekman, J.G., Popa, R.A., Gonzalez, J.E., Stoica, I.:
  Opaque: An oblivious and encrypted distributed analytics platform. In: 14th
  {USENIX} Symposium on Networked Systems Design and Implementation ({NSDI}
  17). pp. 283--298. {USENIX} Association, Boston, MA (2017)

\bibitem{zyskind2015enigma}
Zyskind, G., Nathan, O., Pentland, A.: Enigma: Decentralized computation
  platform with guaranteed privacy (2015),
  \url{https://arxiv.org/pdf/1506.03471.pdf}

\bibitem{Zyskind2015SPW}
Zyskind, G., Nathan, O., Pentland, A.S.: Decentralizing privacy: Using
  blockchain to protect personal data. In: Security and Privacy Workshops
  (SPW). IEEE (2015)

\end{thebibliography}





\end{document}